# Lattice and magnetic structure in the van der Waals antiferromagnet VBr$_3$

Yimeng Gu[1,*], Yiqing Hao[2,*], Zeyu Kao[1,*], Yiqing Gu[1,3], Feiyang Liu[1], Shiyi Zheng[1], Huibo Cao[2], Lunhua He[4,5,6], and Jun Zhao[1,3,7,8†]

[1]State Key Laboratory of Surface Physics and Department of Physics, Fudan University, Shanghai 200433, China

[2]Neutron Scattering Division, Oak Ridge National Laboratory, Oak Ridge, Tennessee 37831, USA

[3]Shanghai Research Center for Quantum Sciences, Shanghai 201315, China

[4]Beijing National Laboratory for Condensed Matter Physics, Institute of Physics, Chinese Academy of Sciences, Beijing, China

[5]Spallation Neutron Source Science Center, Dongguan, China

[6]Songshan Lake Materials Laboratory, Dongguan 523808, China

[7]Institute of Nanoelectronics and Quantum Computing, Fudan University, Shanghai 200433, China

[8]Shanghai Branch, Hefei National Laboratory, Shanghai 201315, China

*These authors contributed equally to this work.
†E-mail: zhaoj@fudan.edu.cn

**Abstract**

We report a comprehensive investigation of the lattice and magnetic structure in van der Waals antiferromagnet VBr$_3$, characterized by a BiI$_3$-type structure at room temperature. Neutron diffraction experiments were performed on both polycrystalline and single-crystalline VBr$_3$ samples, revealing clear magnetic Bragg peaks emerging below the Néel temperature of $T_N$ = 26.5 K. These magnetic Bragg peaks can be indexed by **k** = (0, 0.5, 1) in hexagonal notation. Our refinement analysis suggests that the antiferromagnetic order in VBr$_3$ manifests as a zigzag structure. Moreover, we observed peak splitting for nuclear Bragg peaks in the *HK*-plane below the structure transition temperature of $T_S$ = 90.4 K, indicating the breaking of 3-fold symmetry within the *ab*-plane.

**Introduction**

Two-dimensional van der Waals magnets have attracted significant interest in fundamental and applied physics [1]. This research field presents intriguing challenges to conventional theories and offers opportunities to explore exotic states of matter. Furthermore, it holds great potential for advancing device applications [2]. The recent discovery of long-range ferromagnetic order in the van der Waals ferromagnet CrI$_3$, down to the monolayer limit [3], has sparked considerable research efforts in related systems [4].

More recently, there has been a growing interest in a distinct class of trihalides based on vanadium ions. In contrast to chromium ions, which exhibit quenched orbital moments, vanadium ions have been experimentally shown to possess relatively substantial orbital moments [5] and exhibit stronger spin-orbital coupling [6, 7], as observed in compound like $VI_3$. Consequently, the magnetism of vanadium-based trihalides is anticipated to exhibit heightened sensitivity compared to their chromium-based counterparts when subjected to external modulations such as strain, pressure, and variations in halogen elements [4, 8-13]. This increased sensitivity opens up possibilities for exploring the influence of these external factors on the magnetic and electronic properties of vanadium-based trihalides, offering fresh perspectives for customizing their functionalities and characteristics.

Interestingly, $VBr_3$ seems to exhibit quite different structural and magnetic ordering properties than $VI_3$. At high temperature, $VBr_3$ displays a honeycomb lattice under a space group of R-3 (No. 148 [14]. In contrast to $VI_3$, which undergoes a two-step structural phase transition at $T_{S1}$ = 79 K and $T_{S2}$ = 32 K, $VBr_3$ only undergoes a lattice structure phase transition at $T_S$ = 90.4 K while cooling [14-17]. Despite being theoretically predicted to maintain ferromagnetic order down to the monolayer limit [18], bulk $VBr_3$ undergoes an antiferromagnetic phase transition at $T_N$ = 26.5 K [14-17], exhibiting behavior distinct from extensively studied ferromagnetic counterparts like $CrI_3$ [3] and $VI_3$ [9]. Until now, investigations into the magnetic properties of $VBr_3$ have primarily been conducted at a macroscopic level, leaving the determination of its microscopic magnetic structure unexplored. In this paper, we present a comprehensive investigation of the structural and magnetic phase transitions in $VBr_3$, utilizing a combination of thermodynamic and neutron diffraction measurements.

**Method**

Our $VBr_3$ single crystals were synthesized using the vapor transport technique. Initially, vanadium powder was utilized to deoxidize the $VBr_3$ obtained from Alfa Aesar, which had previously been reported to contain $VBr_2O$ [14]. The two materials were mixed in a molar ratio of 1:3 and then sealed in a silica tube under vacuum. A horizontal three-temperature-zone furnace was employed for sample growth, with the hot end at 480 °C and the cold end at 420 °C. The duration of the growth process was extended from 4 days, as previously reported [14], to 20 days, allowing for the production of large single crystal samples. Large dark brown plate-like single crystals of $VBr_3$ were collected from the cold end of the tubes. $VBr_3$ samples exhibit a high sensitivity to moisture and therefore necessitate handling within a controlled environment such as a glove box.

The heat capacity and field training susceptibility measurements of $VBr_3$ single crystals were conducted using a 9 Tesla Physical Property Measurement System (PPMS, Quantum Design), and the DC-susceptibility of VBr$_3$ single crystals was measured using a 7 Tesla Magnetic Property Measurement System (MPMS, Quantum Design).

The powder neutron diffraction experiment was performed on the General Purpose Powder Diffractometer (GPPD) at China Spallation Neutron Source (CSNS) [19]. Four grams of polycrystalline sample were used in the experiment. The sample was loaded in a powder can made of titanium-zirconium alloy.

Single-crystal neutron diffraction experiments were performed at the four-circle mode of the DEMAND HB-3A at the High Flux Isotope Reactor (HFIR), Oak Ridge National Laboratory (ORNL) [20]. The single crystal used in the experiment measures around 5 mm in its longest dimension, is 0.5 mm thick, and weighs 19 mg. The wavelength was 1.542 Å with a Si (220) monochromator. The data were reduced through ReTIA software [21]. The magnetic structure was refined using Fullprof Suite [22] by the Rietveld method. The Sarah [23] and Bilbao [24-26] programs were used when performing symmetry analysis through representation method [27] and magnetic space group method [28], respectively.

**Results and discussion**

VBr$_3$ was proposed to have a honeycomb-based BiI$_3$-type structure with a space group R-3 above 100 K [14], as shown in Fig. 1(a-b) A structural phase transition is observed through heat capacity and magnetic susceptibility measurements (Fig. 2(a-b)). These measurements reveal anomalies occurring at a critical temperature of $T_S$ = 90.4 K, consistent with previous reports [14-16].

In our single crystal neutron diffraction experiment at HB-3A, nuclear Bragg peaks of VBr$_3$ were collected at three different temperatures: $T$ = 110 K, $T$ = 35 K, and $T$ = 5 K. At 110 K, the nuclear Bragg peaks could be well indexed to a rhombohedral lattice. The lattice parameters (in hexagonal notation) are $a = b$ = 6.408(8) Å and $c$ = 18.45(2) Å. This result is consistent with previous works [14-17] that identified with an R-3 (No. 148) space group at room temperature. As the temperature was lowered below $T_S$ ~ 90 K, we notice an abrupt splitting along the longitudinal direction of the (2, 2, 0) Bragg peak (Fig. 2(c)). This observation suggests a structure phase transition that breaks the C3 symmetry. Consequently, below $T_s$, we encountered three distinct twinned crystal domains characterized by lower symmetry. No significant difference in the crystal structure was observed between 35 K and 5 K.

We attempted to fit the split peak positions to three different lattices: rhombohedral (in hexagonal notation), monoclinic, and triclinic. Our analysis revealed that both the monoclinic and triclinic lattices exhibited comparable fitting qualities for the peak positions, both of which were superior to the rhombohedral lattice. The lattice parameters at 5 K are $a$ = 11.05(3) Å, $b$ = 6.43(3) Å, $c$ = 7.13(4) Å and $\beta$ = 121.1(2) ° in monoclinic cell defined according to Eq. (1). The ratio of $a/b$ in monoclinic notation is 1.718, roughly 1% smaller than $a/b = \sqrt{3}$ in the ideal hexagonal lattice, indicating the breaking of three-fold symmetry across $T_S$. Nevertheless, symmetry analysis shows that breaking of the three-fold symmetry for space group R-3 (No. 148) structure leads to a triclinic P-1 (No.2) structure rather than a monoclinic space group. This is because the parent R-3 structure lacks a two-fold axis, thus a monoclinic lattice cannot be properly defined. As is shown in the isostructural compound VI$_3$, a monoclinic cell can only be defined by using the vacancy-averaged triangular lattice instead of the vacancy-ordered honeycomb lattice [9, 10]. Therefore, a triclinic cell is more appropriate to describe the low temperature structure. The lattice parameters at 5 K are a = 7.13(4) Å, b = 7.17(5) Å, c = 7.11(3) Å, $\alpha$ = 52.87(5) °, $\beta$ = 53.10(2) ° and $\gamma$ = 53.48(6) ° for the triclinic cell defined according to Eq. (2).

$$(a_m, b_m, c_m) = (a_h, b_h, c_h) \begin{pmatrix} 1 & 1 & -1/3 \\ -1 & 1 & 1/3 \\ 0 & 0 & 1/3 \end{pmatrix} \tag{1}$$

$$(a_{tri}, b_{tri}, c_{tri}) = (a_h, b_h, c_h) \begin{pmatrix} 1/3 & 2/3 & -1/3 \\ -2/3 & 1/3 & 1/3 \\ 1/3 & 1/3 & 1/3 \end{pmatrix} \tag{2}$$

Fig. 2(f-g) shows powder diffraction patterns collected at GPPD at 100 K and 35 K. Bragg peaks of V$_2$O$_3$ could be observed in the diffraction pattern as the impurity. We observed an apparent splitting of nuclear peaks with non-zero in-plane components in the data, which is indicative of the disruption of the inherent 3-fold symmetry within the $ab$-plane. As shown in fig. 2(g), the diffraction pattern collected below $T_S$ could be well described by the triclinic cell introduced in Eq. (2). However, it's worth noting that due to the relatively modest neutron scattering cross-section associated with vanadium atoms, we encountered limitations in our ability to fully elucidate the crystal structure of VBr$_3$ at temperatures below $T_S$.

To further characterize the magnetic order of VBr$_3$, we conducted heat capacity and magnetic susceptibility measurements at low temperatures. A clear magnetic phase transition could be observed both in heat capacity and magnetic susceptibility data at $T_N$ = 26.5 K (Fig. 3(e-f)), consistent with previous reports [14-17]. Furthermore, we observed a reduction in $T_N$ by 2.3 K, shifting from 26.5 K to 24.2 K when subjected to a magnetic field of 9 Tesla.

Upon cooling the sample below $T_N$ = 26.5 K, additional peaks emerged in the single-crystal neutron diffraction pattern. These extra peaks could be effectively indexed using a propagation vector of $k_h$ = (0, 0.5, 1) under hexagonal notation. According to the symmetry analysis done by the Bilbao crystallographic server and Sarah software, the magnetic ground state of VBr$_3$ would be either a stripy or a zigzag structure in a magnetic space group of Ps-1 given by parent space group R-3 and $k_h$ = (0, 0.5, 1). However, no magnetic peaks were observed at positions like (0, 1.5, 0) and equivalent positions, effectively ruling out the possibility of a stripy magnetic structure and thereby confirming the presence of the zigzag structure, which is consistent with the magnetic ground state predicted by Dávid et al. through first-principle method [17]. Same results are found if using the actual low temperature space group P-1 and the corresponding propagation vector $k_{tri}$ = (0.5, 0.5, 0). Fig. 3(g) shows the peak intensity measured against temperature at the magnetic Bragg peak (-0.5, 0.5, -1) (under hexagonal natation), revealing the same phase transition temperature as heat capacity and magnetic susceptibility measurements.

It is worth noting that below the structural phase transition, the sample exhibits a division into three domains. Simultaneously, the observed magnetic peaks can be indexed using equivalent k vectors in different domains. To account for these domains, we applied distinct scale factors to each magnetic domain while maintaining the same k-vector during the magnetic structure refinement. The scale factors for the three magnetic domains match those obtained from the corresponding structural domains. The consistency in the relative intensity of magnetic peaks and their correlation with the nuclear domains further supports the conclusion that the magnetic ground state of VBr$_3$ exhibits a zigzag structure with a single propagation vector, rather than the triple propagation vector case proposed in some honeycomb-lattice magnets like Na$_2$Co$_2$TeO$_6$ [29]. Our refinement yields an ordered moment of vanadium ions as follows: $M_x$ = 0.51(5) μ$_B$, $M_y$ = -0.08(4) μ$_B$, and $M_z$ = 0.69(3) μ$_B$, resulting in a total moment of $M_{tot}$ = 0.89(1) μ$_B$. Notably, this total moment is significantly smaller than the classic estimation of pure spin moment of 2$\mu_B$/V$^{3+}$. This deviation can potentially be attributed to the presence of frustrated magnetic interactions within the material or the influence of orbital moments, suggesting a complex interplay of factors influencing the magnetic properties of VBr$_3$. The non-zero M$_y$ case here could not be ruled out assuming the rhombohedral stacking-vacancies are kept at low temperature, similar to the sister compound VI$_3$ [9, 10].

We also attempted to partially-detwin a VBr$_3$ single crystal through the application of an in-plane magnetic field to distinguish the single propagation vector case we proposed here to the multi-**q** case [30]. We conducted susceptibility measurements while applying an external field along the (1, 1, 0) direction, which represents the in-plane projection of one of the easy axes among the three equivalent directions. Our experimental procedure involved initially performing a zero-field cooling measurement with a detection field of 0.05 T. Subsequently, we lowered the temperature of the sample from room temperature which is well above the structural phase transition temperature while subjecting it to a magnetic field of 9 T. Once we reached the base temperature, we reduced the field back to 0.05 T for another field cooling measurement. The collected dataset before and after the field training is presented in Figure 4(d). Analysis of the data revealed that applying a rather small in-plane magnetic field compared to the polarization field of 28 Tesla [17] along the in-plane projection of the easy axis did facilitate the realignment of the three domains associated with the zigzag magnetic structure. It's important to note, however, that while the 9 Tesla in-plane field had an effect, it was not sufficient to achieve complete polarization of the in-plane moments.

It is interesting to compare vanadium-based trihalides with their extensively studied chromium-based counterparts. Both $CrI_3$ [3, 31] and $CrBr_3$ [32, 33] show ferromagnetic order, suggesting a prevailing trend of ferromagnetic interactions. Conversely, $VI_3$ exhibits a ferromagnetic order [9], while $VBr_3$ shows a distinct zigzag-type antiferromagnetic order. The differences in the magnetic interactions between these two groups of compounds may be related to the substantial spin–orbit coupling associated with vanadium-based trihalides. This introduces an additional degree of freedom within the material family, setting vanadium-based trihalides apart from their chromium-based counterparts. To gain a deeper understanding of the magnetic interactions underlying the zigzag-type antiferromagnetic structure in $VBr_3$, further investigations, such as inelastic neutron scattering, are essential. These measurements will help elucidate the precise mechanisms responsible for the zigzag-type magnetic structure similar to those found in Ruthenium-based trihalides [34, 35], as well as the reduced ordered moments in $VBr_3$.

**Conclusion**

In summary, we report a comprehensive examination of the lattice and magnetic structure in the van der Waals antiferromagnet $VBr_3$. Below the Néel temperature ($T_N$ = 26.5 K), we have identified distinct magnetic Bragg peaks that could be successfully indexed as ***k*** = (0, 0.5, 1) in hexagonal notation, indicating a zigzag magnetic structure in $VBr_3$. Furthermore, we have observed the splitting of nuclear Bragg peaks in the *HK*-plane, providing clear evidence of the disruption of 3-fold symmetry within the *ab*-plane of the crystal. These findings form the foundation for a comprehensive understanding of structural and magnetic properties of $VBr_3$.

**Acknowledgments**

This work was supported by the Key Program of the National Natural Science Foundation of China (Grant No. 12234006), the Innovation Program for Quantum Science and Technology (Grant No. 2024ZD0300103), the National Key R&D Program of China (Grant No. 2022YFA1403202) and the Shanghai Municipal Science and Technology Major Project (Grant No. 2019SHZDZX01). This research used resources at the High Flux Isotope

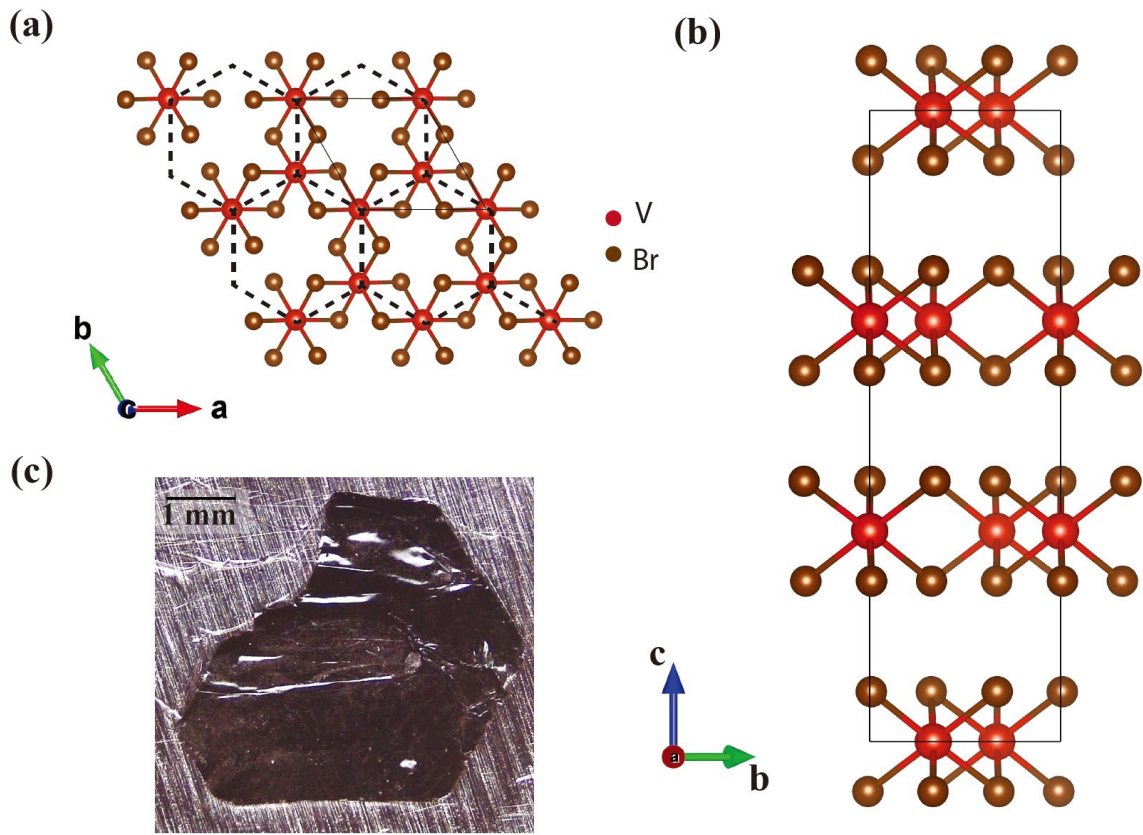

Fig 1. Crystal structure of VBr$_3$. | Crystal structure of VBr$_3$ observed along (a) the *c*-axis and (b) the *a*-axis at 300 K. (c) A photo of a typical dark brown plate-like VBr$_3$ single crystal sample.

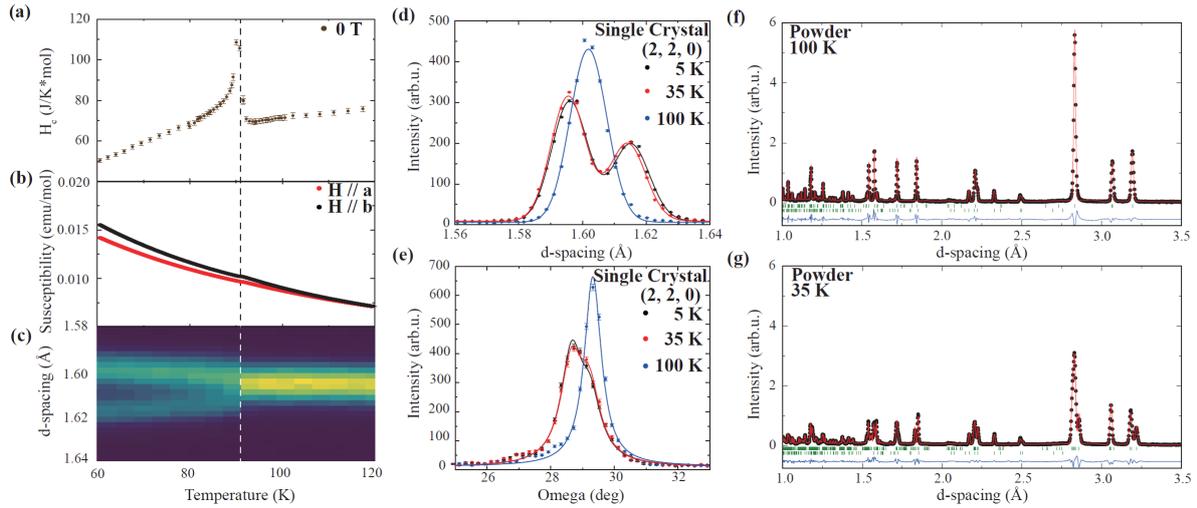

Fig 2. Structure phase transition in VBr$_3$. | (a) Heat capacity and (b) magnetic susceptibility data of VBr$_3$ from 60 K to 120 K. (c) Temperature dependence of nuclear Bragg peak (2, 2, 0). Dashed line indicates the structure phase transition at $T_S$ = 90.4 K. (d) D-spacing splitting and (e) rocking scan of nuclear Bragg peak (2, 2, 0) collected on a VBr$_3$ single crystal sample at HB-3A. Measurements were taken at 5 K, 35 K and 100 K. (f) Powder neutron diffraction pattern of VBr$_3$ at 100 K refined by rhombohedral cell (in hexagonal notation). (g) Powder neutron diffraction pattern of VBr$_3$ at 35 K refined by triclinic cell. Measurements were taken at GPPD, CSNS. According to our refinement of the powder diffraction pattern, the impurity phase V$_2$O$_3$ takes about 12% of the sample and shows no temperature dependence.

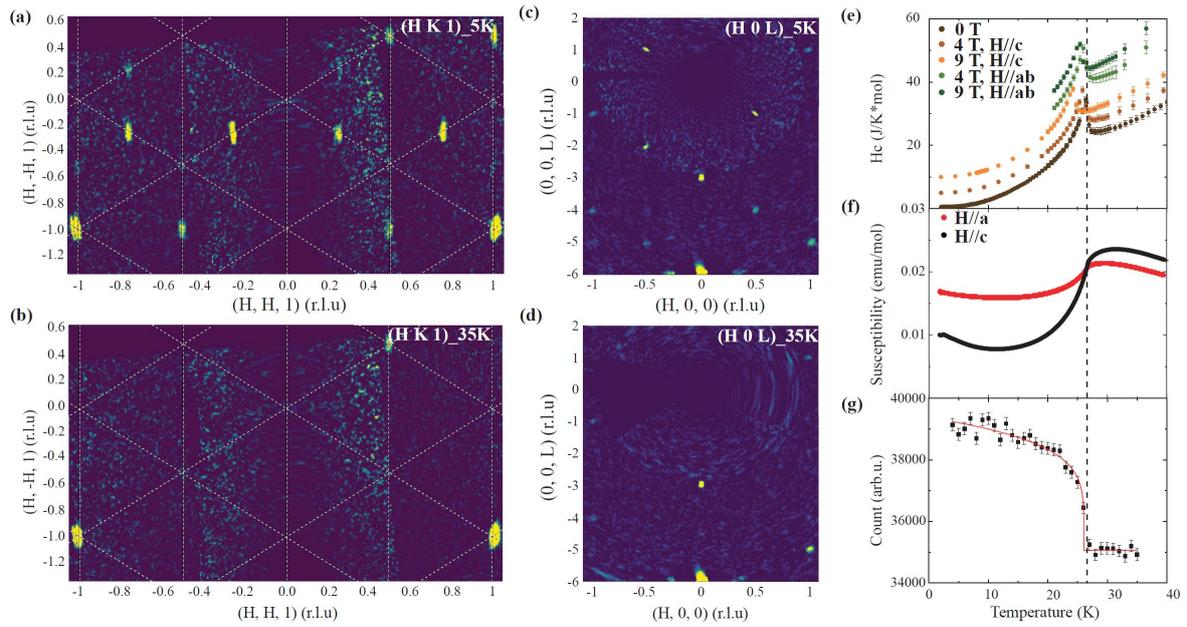

Fig. 3 Magnetic phase transition in VBr$_3$. | Mapping of the reciprocal lattice plane (H, K, 1) of VBr$_3$ at (a) 5 K and (b) 35 K as well as (H, 0, L) plane at (c) 5 K and (d) 35 K collected on VBr$_3$ single crystal sample at HB-3A. The low-temperature part of (e) heat capacity and (f) magnetic susceptibility behavior of VBr$_3$. (g) The temperature dependence of the intensity of magnetic Bragg peak (0.5, -0.5, 1) (under hexagonal notation). Dashed line marks out the magnetic phase transition temperature $T_N$ as 26.06(9) K, consistent with previous reports.

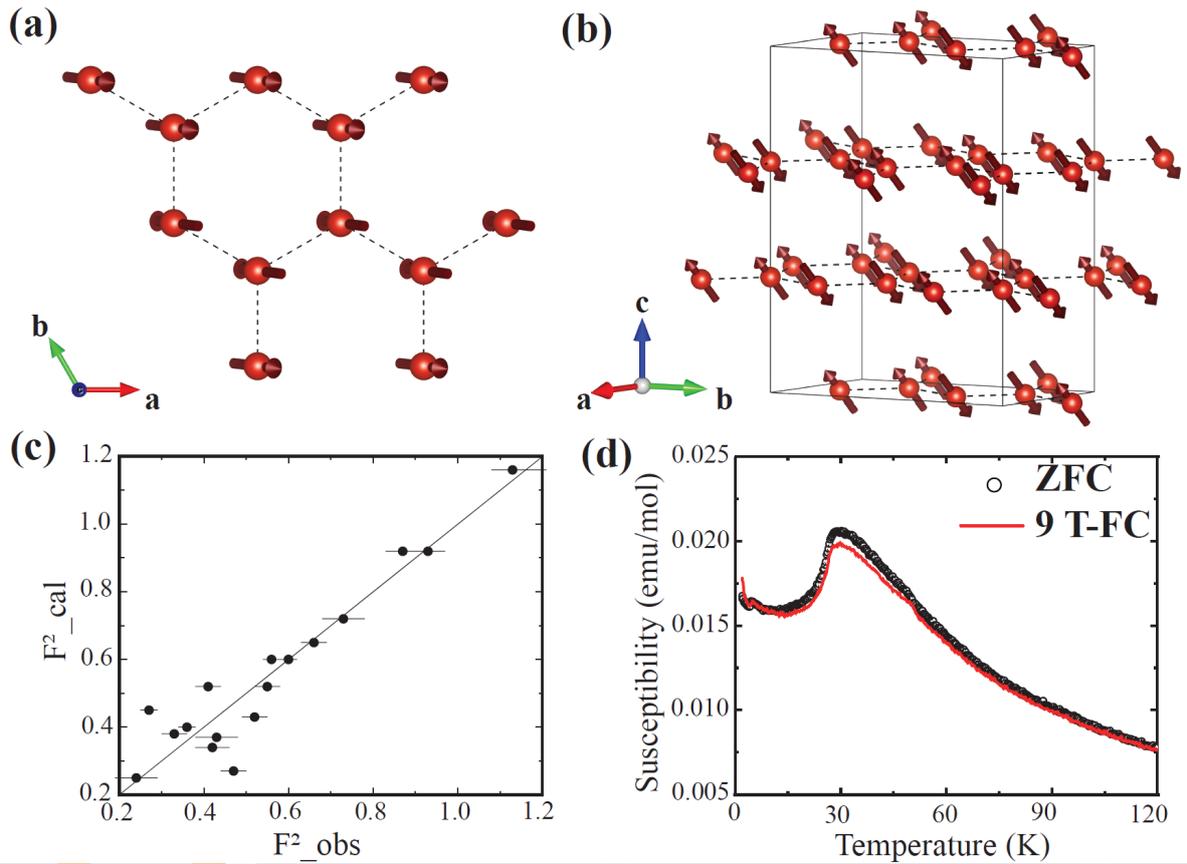

Fig 4. Magnetic ground state of VBr$_3$. | (a) In-plane and (b) overall illustration of zigzag magnetic structure given by the Rietveld refinement. (c) Comparison between measured magnetic structure factor and simulation according to the zigzag magnetic structure model. This magnetic structure model gives a magnetic RF factor of 6.10 and could reproduce the extinction of peaks at (0, 1.5, 0) as well as its equivalent positions. Refinement gives an ordered moment as $M_x$ = 0.51(5) $\mu_B$, $M_y$ = -0.08(4) $\mu_B$ and $M_z$ = 0.69(3) $\mu_B$ with a total moment of $M_{tot}$ = 0.89(1) $\mu_B$. (d) Comparison between zero field cooling magnetic susceptibility and 9 T field cooling magnetic susceptibility of VBr$_3$ measured with external field applied along (1, 1, 0) direction, which is almost parallel to one of the in-plane projection of easy axes for the multidomain zigzag antiferromagnetic structure.